\documentclass[aps,prl,preprint,superscriptaddress,amssymb,amsmath,nofootinbib]{revtex4}

\usepackage{color}
\usepackage{graphicx}
\usepackage{epstopdf}
\usepackage{hyperref} 

\begin{document}

\preprint{}

\title{New Vector Boson Near the Z-pole 
and the Puzzle in Precision Electroweak Data }

\author{Radovan Derm\' \i\v sek}

\affiliation{Physics Department, Indiana University, Bloomington, IN 47405, USA}
\affiliation{Korea Institute for Advanced Study,
Hoegiro 87, Dongdaemun-gu,
Seoul 130-722, Korea}

\author{Sung-Gi Kim}
\affiliation{Physics Department, Indiana University, Bloomington, IN 47405, USA}

\author{Aditi Raval}
\affiliation{Physics Department, Indiana University, Bloomington, IN 47405, USA}

%\email[]{dermisek@ias.edu}

        %\thanks{}
%\altaffiliation {}

%\date{\today}
\date{May 3, 2011}

\begin{abstract}

We show that a $Z'$ with suppressed couplings to the electron compared to the Z-boson,  with couplings to the b-quark, and with a mass close to the mass of the Z-boson, provides an excellent fit to forward-backward asymmetry of the b-quark and  $R_b$ measured on the Z-pole and $\pm 2$ GeV off the Z-pole, and to $A_e$ obtained from the measurement of left-right asymmetry for hadronic final states.  It also leads to a significant improvement in the total hadronic cross section on the Z-pole and $R_b$ measured at energies above the Z-pole. In addition, with a proper mass, it can explain the excess of $Zb\bar b$ events  at LEP in the $90-105$ GeV region of the $b\bar b$ invariant mass.

\end{abstract}

\pacs{}
\keywords{}

\maketitle

%\section{ \label{sec:}}

%\subsection{}

%\subsubsection{}

%\subsection{Introduction}

%\subsubsection{Experimental results}

{\bf Introduction.}
Precision electroweak measurements at LEP, SLC and the Tevatron confirmed numerous predictions of the standard model (SM) with a large degree of accuracy~\cite{:2005ema,Alcaraz:2006mx,Nakamura:2010zzi}. 
Occasionally, deviations from SM expectations appeared, and are still appearing at the Tevatron, however most of them disappeared with more data. Among those that remain, perhaps the longest lasting one,  is a discrepancy in the determination of the weak mixing angle from the LEP measurement of the forward-backward asymmetry of the b-quark,  $A^b_{FB}$,  
%and from  electron asymmetry, $A_e$, obtained from the SLD measurement of left-right asymmetry for hadronic final states.
and from  the SLD measurement of left-right asymmetry for hadronic final states, $A_e{(\rm LR-had.)}$.

These two measurements, showing the  largest deviations from SM predictions among Z-pole observables,  create a very puzzling situation~\cite{Chanowitz:2002cd},\cite{Nakamura:2010zzi}. 
Varying  SM input parameters, especially the Higgs boson mass, one can fit the experimental value for one of them only at the expense of increasing the discrepancy in the other one. While $A^b_{FB}$ prefers a heavy Higgs boson, $m_h \simeq 400$ GeV,  $A_e{(\rm LR-had.)}$ prefers  $m_h \simeq 40$ GeV. Since other observables also prefer a lighter Higgs the focus has been on possible new physics effects that modify $A^b_{FB}$. However, if the pull for a large Higgs mass from $A^b_{FB}$ is removed, the global fit preference 
%for the Higgs boson mass 
%(dominated by $A_e$)  
is in tension with LEP exclusion limit,  $m_h > 114$ GeV~\cite{Barate:2003sz}. 
In addition, it seems difficult to completely explain these deviations by a new physics 
%without ruining the agreement in other observables 
and thus it is widely believed that at least part of the problem is experimental.

We show that a $Z'$ with a mass close to the mass of the Z-boson, provides an excellent fit to measurements of $A^b_{FB}$ on and near the Z-pole,  and simultaneously to $A_e{(\rm LR-had.)}$.
It also improves on the total hadronic cross section on the Z-pole and $R_b$ measured at energies above the Z-pole. In addition, with a proper mass, it can explain the $2.3\sigma$ excess of $Zb\bar b$ events  at LEP in the $90-105$ GeV region of the $b\bar b$ invariant mass.

{\bf $ \bf  Z'$ model.}
We consider a new vector boson, $Z'$, associated with a new gauge symmetry $U(1)'$, with couplings 
 to the electron and the b-quark:%\footnote{To set notation and for later comparison,  in the case of the $Z$ boson, the couplings of a fermion $f$ are given by: $g^f_L =  (T_{3L}^f  - Q^f \sin^2 \theta_W) (- g/\cos \theta_W)$, $g^f_R =  - Q^f \sin^2 \theta_W  (- g/\cos \theta_W)$, where $T_{3L}^f$ is the weak isospin of $f$, $Q^f$ is the electric charge of  $f$, g is the $SU(2)_L$ coupling, and  $ \theta_W$ is the weak mixing angle.}
\vspace{-0.15cm}
\begin{equation}
{\cal L} \supset    Z'_\mu  \bar e \gamma^\mu (g'^e_L P_L  + g'^e_R P_R ) e +
  Z'_\mu \bar b \gamma^\mu (g'^b_L P_L  + g'^b_R P_R ) b.
 \vspace{-0.1cm}
 \nonumber
%g^e_R Z'_\mu \bar e \gamma^\mu P_R e +
%+ h.c.
%\label{eq:lagrangian}
\end{equation}
Without any assumptions about the origin of the $Z'$, all four couplings and the mass of the $Z'$ are treated as free parameters~\cite{Langacker:2008yv}. 
Couplings to other SM fermions and the mixing with the $Z$ boson are assumed to be negligible and are set to zero for simplicity.  Problems associated with a general set of couplings can be cured:
chiral gauge anomalies can be canceled by introducing additional fermions, and Yukawa couplings  can be generated by a  Froggatt-Nielsen type mechanism; or they can be avoided by charging the SM fields under the $U(1)'$ through effective higher dimension operators~\cite{Fox:2011qd}.

{\bf Z' near the Z-pole.}
To demonstrate the basic feature of the effect a Z' can have on precision electroweak data,  let's write the formulas for relevant observables in terms of ``helicity cross  section factors". The differential cross section for $e_L \bar e_L \to f_L \bar f_L$ due to an s-channel exchange of a vector boson is given by  $d \sigma_{LL} / d \cos \theta \propto (g_L^e g_L^f)^2 (1+ \cos \theta)^2$,
where $g^f_{L,R}$ are couplings of the corresponding fermion to the vector boson, and similarly for other helicity combinations: LR, RL, and RR (with a minus sign in front of $\cos \theta$ in the case of LR and RL)~\cite{:2005ema}.
%\footnote{The first index on the cross section corresponds to the helicity of the initial-state electron and the second index corresponds to the helicity of the final-state fermion.} 
Depending on observable, differential cross sections are integrated over various ranges of the scattering angle $\theta$ and thus it is useful to define the helicity cross  section factors as factors in differential cross sections that do not depend on the scattering angle,  $\hat \sigma_{LL} \propto (g_L^e g_L^f)^2$, and similarly for other helicity combinations. In terms of these helicity cross section factors, the forward-backward asymmetry of the b-quark can be written as:
\begin{equation}
A^b_{FB} 
%\equiv \frac{\sigma_F - \sigma_B}{\sigma_F + \sigma_B} 
= \frac{3}{4} \frac{\hat \sigma^b_{LL} -  {\color{blue} \hat \sigma^b_{LR} } -   \hat \sigma^b_{RL} +   \hat \sigma^b_{RR}}{\hat \sigma^b_{LL} +  \hat \sigma^b_{LR} +   \hat \sigma^b_{RL} +   \hat \sigma^b_{RR}} \xrightarrow{Z \; only}  \frac{3}{4} A_e A_b,
\label{eq:AFB}
\end{equation}
where the first part  directly follows from integration of differential cross sections over forward and backward hemispheres. In the case of the Z-boson exchange only the $A^b_{FB}$ reduces to the product of the electron and b-quark asymmetry parameters, defined as $A_f = (g_L^{f2} - g_R^{f2})/(g_L^{f2} + g_R^{f2})$ for a fermion $f$. Similarly, the left-right asymmetry for the b-quark final state can be written as:
\begin{equation}
A^b_{LR} 
%\equiv \frac{\sigma_L - \sigma_R}{\sigma_L + \sigma_R} 
=  \frac{\hat \sigma^b_{LL} +  {\color{blue} \hat \sigma^b_{LR} } -   \hat \sigma^b_{RL} -   \hat \sigma^b_{RR}}{\hat \sigma^b_{LL} +  \hat \sigma^b_{LR} +   \hat \sigma^b_{RL} +   \hat \sigma^b_{RR}} \xrightarrow{Z \; only} A_e,
\label{eq:ALR}
\end{equation}
and in the case of the Z-boson contribution only, it reduces to $A_e$, for any final state. 
The left-right forward-backward asymmetry of the b-quark can be written as:
\begin{equation}
A_{LRFB}^b  = \frac{3}{4} \frac{\hat \sigma^b_{LL} -  {\color{blue} \hat \sigma^b_{LR} } +  \hat \sigma^b_{RL} -   \hat \sigma^b_{RR}}{\hat \sigma^b_{LL} +  \hat \sigma^b_{LR} +   \hat \sigma^b_{RL} +   \hat \sigma^b_{RR}}  \xrightarrow{Z \; only}  \frac{3}{4} A_b,
\label{eq:ALRFB}
\end{equation}
and it is given by $A_b$ in the case of the Z-boson contribution only.
Finally, the ratio of the b-quark and hadronic cross sections can be written as: 
%$R_b  =  (\hat \sigma^b_{LL} +  {\color{blue} \hat \sigma^b_{LR} } +  \hat \sigma^b_{RL} +   \hat \sigma^b_{RR})/{\sum\limits_f  (\hat \sigma^f_{LL} +  \hat \sigma^f_{LR} +   \hat \sigma^f_{RL} +   \hat \sigma^f_{RR})}$.
\begin{equation}
R_b  =  \frac{\hat \sigma^b_{LL} +  {\color{blue} \hat \sigma^b_{LR} } +  \hat \sigma^b_{RL} +   \hat \sigma^b_{RR}}{\displaystyle\sum\limits_f (\hat \sigma^f_{LL} +  \hat \sigma^f_{LR} +   \hat \sigma^f_{RL} +   \hat \sigma^f_{RR})}.
\vspace{-0.3cm}
\label{eq:Rb}
\end{equation}

%Over the years the the discrepancies in Z-pole observables related to the b-quark varied, starting from ``$R_b$-$R_c$" puzzle with $R_b$ measured value exceeding the standard model prediction by $\sim 3\sigma$ to the current stage when $R_b$ agrees with the SM prediction quite precisely and $A_{FB}^b$ disagrees at $2.5 \sigma$. Variety of solutions were suggested. We will argue that the for the  $Z'$ near the Z-pole is a unique possibility providing simultaneous  explanation of many measurements. 

Previous explanations of the deviation in  $A^b_{FB}$ focused on modifying $g_R^b$. 
Achieving this and simultaneously not upsetting quite precise agreement in $R_b$ turned out to be very challenging for a new physics that enters through loop corrections~\cite{Haber:1999zh}. This motivated  tree level modification of the $g_R^b$ either through mixing of b-quark with extra fermions~\cite{Choudhury:2001hs} or through Z-Z' mixing~\cite{He:2002ha}. 
However the $A^b_{FB}$ is only a part of the puzzle and, as is clear from Eq.~(\ref{eq:ALR}),  any new physics that reduces to modification of bottom quark couplings cannot affect the $A_e{(\rm LR-had.)}$.
%$A_e$ obtained from left-righ asymmetry measurements.

We suggest to modify the $b \bar b$ production cross section directly, $e^+e^- \to Z'^* \to b \bar b$, rather than modifying the Z-couplings. 
This idea comes from a simple observation that increasing ${\color{blue} \hat \sigma^b_{LR} }$ so that $R_b$ increases by about 0.4\% (which still produces a better fit than the standard model) decreases $A^b_{FB}$, see Eq. (\ref{eq:AFB}),  by $\sim 4\%$ which is exactly what is needed to fit the experimental value. This 10 times larger effect is a result of an approximate, $\sim  90\%$, cancellation between  $\hat \sigma^b_{LL}$ and $ \hat \sigma^b_{RL}$ in the SM due to comparable $g^e_L$ and $g_R^e$ couplings. 
 For the same reason, $A_e{(\rm LR-had.)}$ increases by $\sim 4\%/5 = 0.8\%$, see Eq. (\ref{eq:ALR}) (the factor of 5 comes from  $b \bar b$ representing $\sim 20\%$ of hadronic final states), which brings it to $\sim 1\sigma$ from the experimental value.
 %, see Table~\ref{tab:fit}.

To generate a sizable contribution to $A^b_{FB}$ on the Z-pole and not significantly affect predictions for $A^b_{FB}$ and $R_b$ above the Z-pole (that roughly agree with measurements), the increase in ${\color{blue} \hat \sigma^b_{LR} }$ must be due to  s-channel exchange of a new vector particle with mass close to the mass of the Z-boson. A scalar particle near the Z-pole can modify $A^b_{FB}$  only comparably to its modification of $R_b$. This was considered in Ref.~\cite{Erler:1996ww} motivated by previous discrepancies in Z-pole observables. Similarly $Z'$ was used to explain previous discrepancies, see {\it e.g.} a heavy $Z'$~\cite{Erler:1999nx} or almost degenerate $Z$ and $Z'$~\cite{Caravaglios:1994jz} scenarios.
A heavy  particle, or a particle contributing in t-channel, can modify Z-pole observables only negligibly if it should not dramatically alter predictions above the Z-pole.
Thus a $Z'$ near the Z-pole with small couplings to the electron (in order to satisfy limits from searches for $Z'$) and sizable couplings to the bottom quark is the only candidate.

{\bf Numerical analysis.}
We construct a $\chi^2$ function of relevant quantities related to the bottom quark and electron measured at and near the Z-pole which are summarized in Table~\ref{tab:fit}. Their precise definition can be found in the EWWG review~\cite{:2005ema} from which we also take the corresponding experimental values.
Instead of the pole  forward-backward asymmetry of  the b-quark, $A_{FB}^{0,b}$, we include three measurements of the asymmetry, at the peak and $\pm 2$ GeV from the peak. These are more relevant because the presence of a $Z'$ near the $Z$-pole changes the energy dependence of the asymmetry. In addition,  about 25\% of the deviation in the pole asymmetry comes from the measurement at $+2$ GeV from the peak. Corresponding LEP averages for $R_b$ at  $\pm 2$ GeV from the peak do not exist. These are available only from DELPHI~\cite{Abreu:1998xf} and although they are included in the Z-pole LEP average,  $R_b^0$, we include them in addition in order to constrain the energy dependence.
 We further  include pole values of the total hadronic cross section, $\sigma^0_{\rm had}$, the ratio of the hadronic and electron decay widths, $R_e^0$, forward-backward asymmetry of  the electron, $A_{FB}^{0,e}$, measured at LEP; and  SLD values of asymmetry parameters of the b-quark,   $A_b$, obtained from the measurement of left-right forward-backward asymmetry, and the electron, obtained from the measurement of left-right asymmetry for hadronic final states, $A_e{(\rm LR-had.)}$, and leptonic final states, $A_e{(\rm LR-lept.)}$.

We calculate theoretical predictions using ZFITTER 6.43~\cite{Bardin:1999yd,Arbuzov:2005ma} and  ZEFIT 6.10~\cite{Leike:1991if} which we modified for a $Z'$ with free couplings to the b-quark and the electron.  
%In the case of the standard model we precisely reproduce the result in the EWWG review~\cite{:2005ema} or the PDG review~\cite{Nakamura:2010zzi} for sets of SM input parameters used in those fits. 
In our fit we use the SM input parameters summarized in Table 8.1 of the EWWG review~\cite{:2005ema}, namely: $m_Z = 91.1875$ GeV, $\Delta \alpha^{(5)}(m_Z^2) = 0.02758$, $\alpha_S(m_Z^2) = 0.118$; however, we update the  top quark mass  to the Tevatron average, $m_t = 173.3$ GeV~\cite{:1900yx}, and fix the Higgs mass to $m_H = 117$ GeV. 
%{\color{green}{{As a result of the different set of SM input parameters, our SM predictions, given in Table~\ref{tab:fit}, are slightly different from~\cite{:2005ema,Nakamura:2010zzi}. The effects of varying input parameters on electroweak observables can be found in~\cite{:2005ema}. The differences resulting from a given choice of SM input parameters are not essential for comparison of the SM and SM+$Z'$.}}}
We minimize the $\chi^2$ function of 5 parameters, $m_{Z'}$, $g'^e_L$, $g'^e_R$, $g'^b_L$, and $g'^b_R$, with MINUIT~\cite{minuit}. In principle, the width,  $\Gamma_{Z'}$, could be treated  as a free parameter because $Z'$ can have additional couplings 
%besides those given in Eq.~(\ref{eq:lagrangian}) 
that do not affect precision electroweak data. For simplicity, we do not consider this possibility.

\begin{figure}[t] %  figure placement: here, top, bottom, or page
   \centering
     \includegraphics[width=3.in]{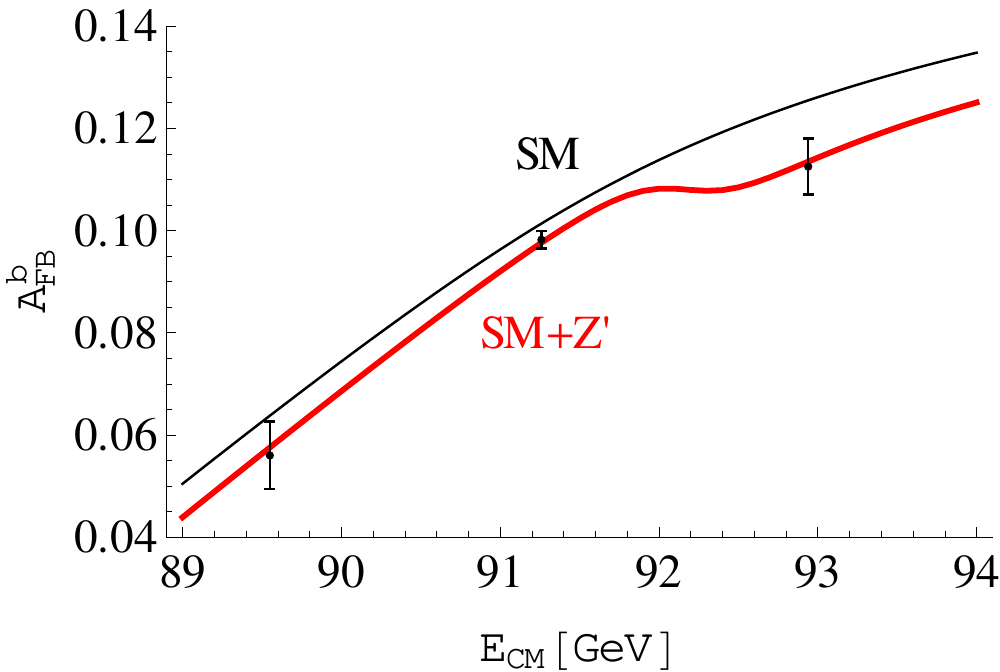}      
   \includegraphics[width=3.in]{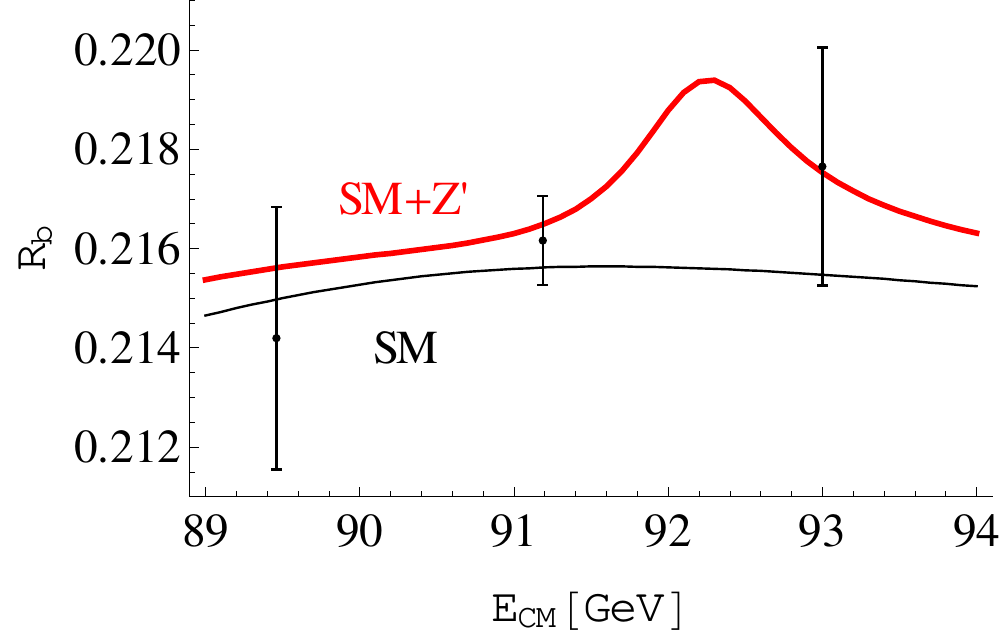}  
   \includegraphics[width=3.in]{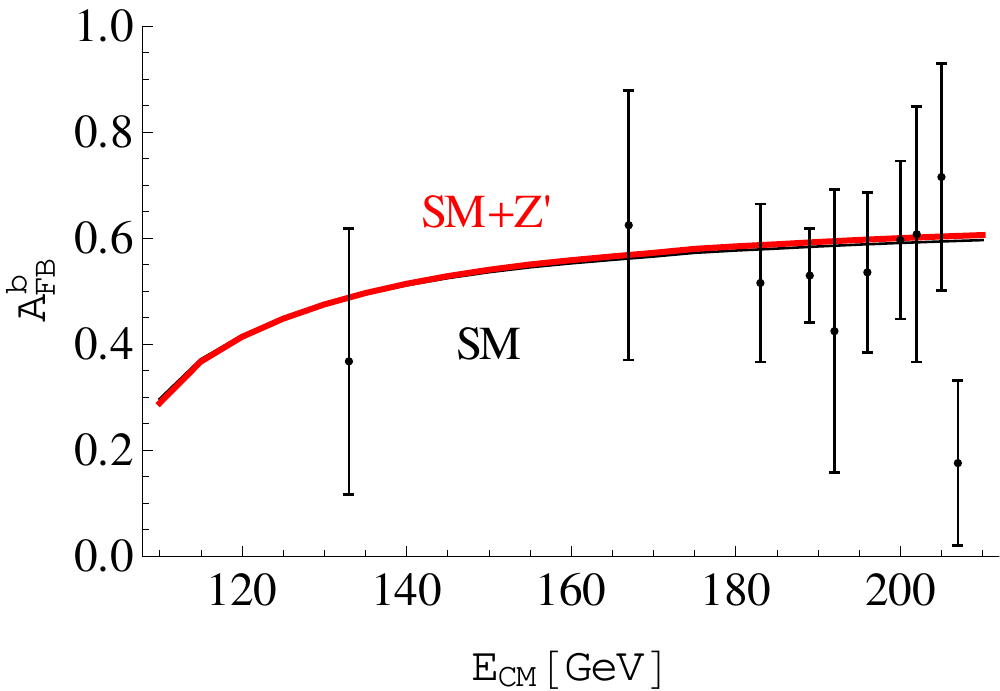} 
        \includegraphics[width=3.in]{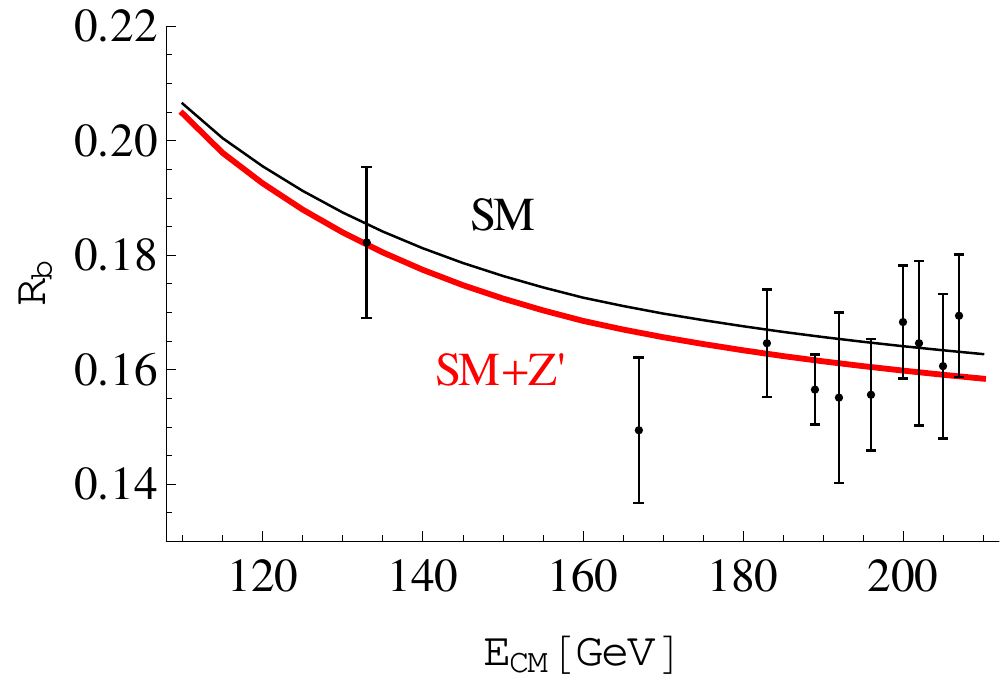} 
   \caption{Experimental values of $A_{FB}^b$ and $R_b$ and predictions of  the SM (thin lines) and the Z' model (thick lines) for  input parameters specified in the caption of Table~\ref{tab:fit} as functions of center of mass energy near and above the Z-pole.}
   \label{fig:NP}
\end{figure}

{\bf The best fit solution.} 
The best fit to precision data included in the $\chi^2$ is summarized in  Table~\ref{tab:fit} and  parameters for which the best fit is obtained are given in the caption. 
Clearly, addition of $Z'$ provides an excellent fit to selected precision electroweak data with $\chi^2 = 4.6$ for 12 obsevables with 5 additional parameters compared to the standard model that has $\chi^2 = 22$. The most significant improvement comes from the three measurements of $A_{FB}^b$ which can be fit basically at central values in the $Z'$ model, without spoiling the agreement in $R_b$. The energy dependance of both quantities near the Z-pole  for both the SM and $Z'$ model together with data points is plotted in Fig.~\ref{fig:NP}.
% which clearly illustrates the virtue of the $Z'$ model. 
The $A_e{(\rm LR-had.)}$ and $\sigma_{\rm had}^0$ are also fit close to their central values.

%However,  flavor diagonal couplings to other quarks might not be negligible, and couplings to neutrinos and  particles beyond the standard model might be sizable without violating experimental limits. Thus the width of the $Z'$ might also be treated as a free parameter.
%

\begin{table}[ht]
    \caption{The best fit to relevant precision electroweak observables in the SM with a $Z'$.
    The best fit is achieved for: $m_{Z'} = 92.2$ GeV, 
    %$\Gamma_{Z'} = 1.1$ GeV, 
    %and couplings  
    $g'^e_L = 0.0059$, $g'^e_R = 0.0073$, $g'^b_L = 0.040$,  and $g'^b_R = -0.54$; ($\Gamma_{Z'} = 1.1$ GeV).
    The standard model input parameters are fixed to $m_t = 173.3$ GeV, $m_h = 117$ GeV, and other parameters as listed in Table 8.1 of the  EWWG review~\cite{:2005ema}. For comparison, we also include predictions of the standard model with $\chi^2$ contributions.
    % for the same set of fixed input parameters.
    }

\begin{tabular}{  l  c  c c c c }
 \hline
  \hline
Quantity    & Exp. value & SM & $\chi^2_{SM}$ & $Z'$ & $\chi^2_{Z'}$\\
 \hline
     {\color{blue}{$\sigma_{\rm had}^0$   [nb]} }     & 41.541(37) &  41.481 &  {\color{blue}{{\bf 2.6}}}& 41.529 &  {\color{blue}{{\bf 0.1}}}\\
    $R_b (-2)$            & 0.2142(27)		     &  0.2150 & 0.1 & 0.2156 & 0.3 \\
       $R_b^0$            &  0.21629(66)	   &  0.21580 & 0.6 & 0.21670 & 0.4 \\
       $R_b (+2)$            &  0.2177(24)	   & 0.2155 & 0.8 & 0.2177 & 0.0\\
       {\color{blue}{$A_{FB}^b (-2)$  }}          & 0.0560(66)		     &  0.0638 &{\color{blue}{{\bf 1.4}}}& 0.0577 & {\color{blue}{{\bf 0.1}}}\\
       {\color{blue}{$A_{FB}^b ({\rm pk})$  }}          &  0.0982(17)	   &  0.1014  &   {\color{blue}{{\bf 3.5}}} & 0.0979 & {\color{blue}{{\bf 0.0}}} \\
       {\color{blue}{$A_{FB}^b (+2)$   }}         &  0.1125(55)	   &  0.1255 & {\color{blue}{{\bf 5.6}}}& 0.1136 & {\color{blue}{{\bf 0.0}}}\\
      $A_b$       &  0.923(20) &  0.9346 & 0.3 & 0.9237 & 0.0 \\
      $R_e^0$       &  20.804(50) &  20.737 & 1.8 & 20.765 & 0.6 \\
                  $A_{FB}^{0,e}$       &  0.0145(25) &  0.0165  & 0.7 & 0.0174 & 1.4 \\
      {\color{blue}{$A_e{(\rm LR-had.)}$ }}      &  0.15138(216) & 0.14739  & {\color{blue}{{\bf 3.4}}} & 0.15014 & {\color{blue}{{\bf 0.3}}}\\
      $A_e{(\rm LR-lept.)}$      &  0.1544(60) & 0.1473  & 1.4 & 0.1473 & 1.4 \\
        \hline
      total  $\chi^2$ &    &   &  22.1 & & 4.6\\
            \hline
  \hline
      \end{tabular}
   \label{tab:fit}
\end{table}

Besides quantities included in the $\chi^2$ and given in  Table~\ref{tab:fit} we check all other electroweak data on and near the Z-pole,  and above and below the Z-pole. 
While b-quark quantities were measured at three energies near the Z-pole, the total hadronic cross section was measured also at $\pm 1,3$ GeV (from data collected only during 1990-1991). The measurement at $+1$ GeV roughly coincides with the Z'-peak where the deviation from the SM would be the largest. The experimental error in  $\sigma_{\rm had}$  at  $+1$ GeV from the peak is $\sim 1\%$ for each LEP experiment and thus the Z'-peak contributes only a  fraction of the error bar. 
%Similarly the error at $\pm 2$ GeV  is about $0.3\%$ for each experiment. 

At energies above the Z-pole, the  $A_{FB}^b$ in the $Z'$ model basically coincides with the SM prediction while
$R_b$  fits data better than the SM, see Fig.~\ref{fig:NP}, with   $\chi^2 = 4.8$ for 10 data points compared to the SM which has   $\chi^2 = 7.2$ (the average discrepancy with respect to the SM prediction for $R_b$ is $-2.1\sigma$)~\cite{Alcaraz:2006mx}. At energies below the Z-pole the $Z'$ leads only to negligible differences from the SM predictions compared to sensitivities of current experiments.  

%The predicted deviations from standard model predictions at low energy experiments including B-factories are too small compared to sensitivities of these experiments.
%error bars on lepton universality are of order $\gtrsim 2\%$.

%\begin{figure}[htbp] %  figure placement: here, top, bottom, or page
 % \centering
  %\includegraphics[width=1.65in]{AFBbOP_04252011.pdf} 
    % \includegraphics[width=1.65in]{RbOP_04252011.pdf} 
   %\caption{The same as in Fig.~\ref{fig:NP} for energies above the Z-pole.}
   %\label{fig:OP}
%\end{figure}

The quantities related to other charged leptons and quarks are not directly affected by $Z'$ and the predictions are essentially identical to predictions of the SM~\cite{Nakamura:2010zzi}.   
For example, the LEP 1 average of leptonic asymmetry assuming lepton universality, $A_l = 0.1481 \pm 0.0027$,
agrees very well with the SM prediction and would be only negligibly altered by the $Z'$ with couplings corresponding to the best fit (the prediction is  the same as for $A_e{(\rm LR-lept.)}$ given in Table~\ref{tab:fit}).

 {\bf Other fits.} The full exploration of the $Z'$ parameter space is beyond the scope of this letter. However it is instructive to make few comments. The $\chi^2$ is a very shallow function of the $Z'$ parameters, except the $Z'$ mass. Varying couplings by $10\%$ leads to a comparable fit. Actually, almost all the improvement in the  $\chi^2$ comes from the $g'^e_L$ and $g'^b_R$ couplings  because these are needed to modify ${\color{blue} \hat \sigma^b_{LR} }$ as discussed above. With only these two couplings the best fit is achieved for: $m_{Z'} = 92.1$ GeV,  $g'^e_L = 0.0048$, and  $g'^b_R = -0.47$ with $\chi^2 = 6.4$ (only slightly worse than the best fit with all the couplings). In addition, values of couplings separately  are not crucial, as far as $g'^e_L$ is small, not  upsetting  electron observables. Thus this striking improvement in the $\chi^2$ for the Z-pole and near the Z-pole  observables, is achieved with only two relevant parameters: $m_{Z'}$ and the product of couplings, $(g'^e_L g'^b_R)$.
 
 Increasing the $Z'$ mass the fit gets worse, mostly driven by near the Z-pole measurements of $R_b$
 (corresponding near the Z-pole values of the total hadronic cross section, which are not included in the $\chi^2$, are also affected).
 Fixing the  $Z'$ mass to 95 GeV the best fit is achieved for somewhat larger couplings to electrons: $g'^e_L = 0.027$,  $g'^e_R = 0.013$, $g'^b_L = 0.08$, and  $g'^b_R = -0.49$ with $\chi^2 = 9.3$ which is still a significant improvement from the SM. Increasing the $Z'$ mass above $\sim 110$ GeV improves the fit to precision electroweak data only marginally.

{\bf Other constraints.}
At LEP $Z'$ could be produced together with the Z-boson, $e^+e^- \to Z Z'$, or pair produced. If other couplings besides $g_{L,R}^{e,b}$ are absent the $Z'$ would decay to $b \bar b$ with branching ratio close to 100\% and thus  it would result in a small excess in $Z b \bar b$ and a negligible excess in $b \bar b b \bar b$ data that were closely scrutinized  in  searches for  Higgs bosons. The search for the SM Higgs boson in $Zb \bar b$ final state shows a $2.3\sigma$ excess of events for the $b \bar b$ invariant mass in the range $90 -105$ GeV~\cite{Barate:2003sz}. It is compatible with  $\sim 10\%$ of the SM Higgs production cross section for $m_h = 100$ GeV,
 and thus it can be  explained either by a Higgs boson with reduced coupling to the Z-boson~ \cite{Sopczak:2001tk,Carena:2000ks,Drees:2005jg} or
a SM-like Higgs boson with reduced branching fraction to $b \bar b$~\cite{Dermisek:2005ar,Dermisek:2005gg,Dermisek:2007yt}. 

The $Z'$ with properties studied in this paper can provide another explanation.  The best fit presented in Table~\ref{tab:fit} can explain only a fraction of the excess (extra $\sim 5$ fb of $Z b \bar b$ cross section). However, as already mentioned, 
increasing $g'^e_{L,R}$ and decreasing $g'^b_{L,R}$ so that their products are the same leads to small differences in the $\chi^2$. Thus $\sigma(e^+e^- \to Z Z')$ which depends only on $g'^e_{L,R}$ can be adjusted. 
%the $\chi^2$ is a shallow function of couplings especially in the direction of increasing couplings to the electron and decreasing couplings to the b-quark (so that the products are the same) and thus the $e^+e^- \to Z Z'$ can be adjusted leading only to a small difference in $\chi^2$. 
For example, the best fit with fixed $m_{Z'} = 95$ GeV (see above) contributes extra 36 fb of $Z b \bar b$ cross section, which is about $10\%$ of the SM Higgs production cross section, perfectly matching the excess.

The same search also showed a deficit of $Z b \bar b$ events for the $b \bar b$ invariant mass below the Z-mass. 
It would be interesting to see whether this deficit can be a result of the negative interference of $Z'$ with $\gamma$ and $Z$ in $e^+e^- \to Z (\gamma^*, Z^*, Z'^*) \to Z b \bar b$. This requires a careful study.

%{\color{green}{Finally, with $Z \to b \bar b$, the same process would also lead to about $\sim 5.6$ fb of $b  \bar b   b \bar b$ for $m_Z+m_{Z'} \simeq 186$ GeV which is very small compared to limits and could show up only as a small bump  in the search for $hA$ production for $m_h+m_A \simeq 186$ GeV. }}

At the Tevatron this $Z'$ could be produced only in association with b-quarks. The $Z' b$ cross section  can be easily estimated
from studies of the $Z b$ production which is  a background for Higgs searches~\cite{Campbell:2003dd}. For the three fits discussed above we find $\sigma(p \bar p \to Z'b) \simeq 20 - 30$ pb. Both CDF and D0 searched for the Higgs boson produced in association with b-quarks~\cite{CDF_3b,Abazov:2010ci}, 
%The strongest constraint come from the CDF which sets limits on  
and set limits
$\sigma (p \bar p \to Hb) \times B(H\to b \bar b) < 30 - 50$ pb for $m_H \simeq 90-100$ GeV~\cite{CDF_3b}. This is not very far from the prediction and thus updated analyses with larger data sets might see an excess or start constraining the size of $g'^b_R$. 
At the LHC the  $Z' b$ cross section is two orders of magnitude larger~\cite{Campbell:2003dd} and it is just a question of accumulating enough luminosity to see the signal of $Z'$.  
Note however, that with possible couplings of $Z'$ to other quarks  (or particles beyond the SM) the $B(Z' \to b \bar b)$ can be highly reduced which could make the search for $Z'$ difficult. 

%The CDF search for Higgs boson produced in association with b-quarks~\cite{CDF_3b} (with $2.2 {\rm fb}^{-1}$ of integrated luminosity) sets limits on $\sigma (p \bar p \to Hb) \times B(H\to b \bar b) \lesssim 30$ pb for $m_H \simeq 90$ GeV.

%The D0 search for Higgs boson produced in association with b-quarks~\cite{Abazov:2010ci} (with $5.2 {\rm fb}^{-1}$ of integrated luminosity) sets limits on $\sigma (p \bar p \to Hb) \times B(H\to b \bar b) \lesssim 90$ pb for $m_H \simeq 90$ GeV.

{\bf Conclusions and Outlook.} The $Z'$ near the Z-pole with couplings to the electron and the b-quark can resolve the puzzle in precision electroweak data by explaining the two largest deviations from SM predictions among Z-pole observables: $A_{FB}^b$ and  $A_e{(\rm LR-had.)}$. It nicely fits the energy dependence of $A_{FB}^b$ near the Z-pole and improves on  $\sigma_{\rm had}^0$ on the Z-pole and $R_b$ measured at energies above the Z-pole. Certainly it is possible that all these deviations from the SM are just statistical fluctuations and systematic errors, or a combination of these with effects of much more complicated new physics.
However it is intriguing that these deviations, together with the $2.3\sigma$ excess of $Zb\bar b$ events  at LEP that can be fully explained by $Z'$,  might as well be hints of a new force of nature.

Besides the Tevatron and the LHC, where this $Z'$ might be seen in b-quark rich events, the optimal experiment to confirm or rule out this possibility would be the future linear collider, especially the GigaZ option, which would allow more accurate exploration of the Z-peak.  

Considering other flavor conserving couplings, or small flavor violating couplings,  expands the range of observables to which this Z' could contribute. It would be interesting to see if it can simultaneously explain some other deviations from SM predictions.

%%%%%%%%%%%%%%%%%%%%%%%%%%%%%%%%%%%%%%%%%%%%%%%%%%%%%%%%
%\acknowledgements
%%%%%%%%%%%%%%%%%%%%%%%%%%%%%%%%%%%%%%%%%%%%%%%%%%%%%%%%

\vspace{0.2cm}
{\it Acknowledgments:} We thank D. Bourilkov, M. Gruenewald, H.D. Kim, R. Van Kooten, K. Moenig, P. Langacker and members of KIAS for useful discussions.  
RD thanks KIAS and SNU, especially E.J. Chun and H.D. Kim,  for kind hospitality and support during the final stages of this project. This work was supported in part 
by the award from Faculty Research Support Program at Indiana University.

%%%%%%%%%%%%%%%%%%%%%%%%%%%%%%%%%%%%%%%%%%%%%%%%%%%%%%%%

%%%%%%%%%%%%%%%%%%%%%%%%%%%%%%%%%%%%%%%%%%%%%%%%%%%%%%%%%

%%%%%%%%%%%%%%%%%%%%%%%%%%%%%%%%%%%%%%%%%%%%%%%%%%%%%%%%%
\end{document}